\documentclass[osajnl,twocolumn,showpacs,superscriptaddress,11pt,floatfix]{revtex4-1} %% use 11pt for Applied Optics
\usepackage{amsmath,amssymb,graphicx}
\begin{document}

\title{Investigation of the Young's modulus and thermal expansion of amorphous titania-doped tantala films}

\author{Matthew R. Abernathy}\email{Corresponding author: abernathy\_m@ligo.caltech.edu}%
\affiliation{LIGO, Caltech, 1200 E California Blvd, Pasadena, CA 91125, USA}%
\affiliation{SUPA, University of Glasgow, University Ave, Glasgow, G12 8QQ, UK}

\author{James Hough}
\author{Iain W. Martin}
\author{Sheila Rowan}
\affiliation{SUPA, University of Glasgow, University Ave, Glasgow, G12 8QQ, UK}

\author{Michelle Oyen}
\affiliation{Cambridge University, Engineering Department, Trumpington Street, Cambridge, CB2 1PZ, UK}

\author{Courtney Linn}
\affiliation{Embry-Riddle Aeronautical University, 3700 Willow Creek Road, Prescott, AZ 86301, USA}

\author{James E. Faller}
\affiliation{JILA, University of Colorado, 440 UCB, Boulder, CO 80309, USA}

\begin{abstract}
The current generation of advanced gravitational wave detectors utilize titania-doped tantala/silica multilayer stacks for their mirror coatings. The properties of the low-refractive-index silica are well known; however, in the absence of detailed direct measurements, the material parameters of Young's modulus and coefficient of thermal expansion (CTE) of the high refractive index material, titania-doped tantala, have been assumed to be equal to values measured for pure tantala coatings. In order to ascertain the true values necessary for thermal noise calculations, we have undertaken measurements of Young's modulus and CTE through the use of nanoindentation and thermal-bending measurements. The measurements were designed to assess the effects of titania doping concentration and post-deposition heat-treatment on the measured values in order to evaluate the possibility of optimizing material parameters to further improve thermal noise in the detector. Young's modulus measurements on pure tantala and 25\% and 55\% titania-doped tantala show a wide range of values, from 132 to 177 GPa, dependent on both titania concentration and heat-treatment. Measurements of CTE give values of $(3.9\pm0.1)\times10^{-6}\text{ K}^{-1}$ and $(4.9\pm0.3)\times10^{-6}\text{ K}^{-1}$ for 25\% and 55\% titania-doped tantala, respectively, without dependence on post-deposition heat-treatment. %Fortunately, the traditional values used in thermal noise calculations within the LSC, 140 GPa and $(3.6\pm0.1)\times10^{-6}\text{ K}^{-1}$, are very close to the appropriate values measured here.
\end{abstract}

% REPLACE WITH CORRECT OCIS CODES FOR YOUR ARTICLE
                          % NOTE: \ocis{} IS ALIASED TO \pacs{} BUT MUST
                          % FORMAT THE TERMS CORRECTLY FOR EACH JOURNAL

\maketitle %% required
\section{Introduction}
The current generation of interferometric gravitational wave detectors, including the Advanced LIGO \cite{Harry10} and Advanced Virgo \cite{AVIRGO} detectors, are undergoing construction, and are expected to reach design sensitivity in the next few years. An important limiting noise source in the detectors is the thermal noise arising from the coatings used to make the mirrored test masses reflective at a wavelength of 1064 nm. These high-reflectivity coatings are made from alternating layers of high index of refraction Ion Beam Sputtered (IBS) amorphous titania-doped tantala (Ti:Ta$_2$O$_5$) and low index IBS amorphous silica (SiO$_2$), with the layer structure optimized to reduce thermal noise while maintaining the requisite reflectivity \cite{Villar10}. 

In order to calculate the thermal noise that arises in the interferometers a priori, the thermo-mechanical properties of the coating materials need to be well known. In order to calculate the thermo-optic noise \cite{Evans08} of the coatings, knowledge of the heat capacity, CTE, thermo-optic coefficient, thermal conductivity, Young's modulus, and Poisson ratio of the coating material is required. In order to calculate the Brownian thermal noise \cite{Harry02}, the Young's modulus, Poisson ratio, and mechanical loss of the coating material is also needed. Furthermore, calculation of the mechanical loss from various `ring-down' measurement techniques \cite{Penn03,Martin09b} requires knowledge of the coating material's Young's modulus and Poisson ratio.   

While the properties of silica are fairly well known, the properties of tantala, and especially titania-doped tantala, have rarely been measured. In some cases, there has even been some controversy regarding some measurements. In the case of the CTE, measurements have ranged from $-4.4\times10^{-5}$ K$^{-1}$ for ion-assisted e-beam sputtered tantala \cite{Inci04}, to $2.4\times10^{-6}$ K$^{-1}$ for IBS tantala \cite{Tien00}. Measurements made by Braginsky and Samoilenko \cite{Braginsky03a} suggested that the CTE of tantala was roughly $(5\pm1)\times10^{-6}$ K$^{-1}$. This new measurement was used to support the use of a value of $3.6\times10^{-6}$ K$^{-1}$ within the LIGO community for IBS pure tantala and it has since been commonly used \cite[and others]{Evans08,Fejer04}. However, this value was calculated \cite{Tien00} from measurements of the temperature coefficients \cite{Scobey94} and normalized thermo-optic coefficients \cite{Chu97}, and was not a direct measurement of thermal expansion, nor were these values measured on IBS coatings.

For the Young's modulus of tantala, the value of 140 GPa is most often used in analysis of coating mechanical loss and estimates of coating thermal noise. The article by Martin et al. \cite{Martin93} is often cited \cite[and others]{Crooks04,Penn03,Martin08,Martin09b}; however, this paper only displays plots of indentation modulus as a function of indentation depth using a micro-indentation system, and these plots have not been fully analyzed to give an appropriate coating Young's modulus. Other measurements support the value of 140 GPa using nanoindentation, including measurements of $140\pm15$ GPa \cite{Gross04}, and 143 GPa \cite{Cetinorgu09}. Unfortunately, these measurements do not offer a complete analysis of the substrate effects on the nanoindentation measurements.  

The purpose of the measurements made in this paper is twofold: first, to remove any further controversy regarding the Young's modulus and CTE of IBS tantala films; and second, to measure these properties for the case of the IBS titania-doped tantala coatings used in advanced detectors. As dopant level and post-deposition heat treatment have been identified as variables that can affect the Brownian thermal noise in the detectors \cite{Harry07,Martin10,Martin09b}, coatings with a range of dopant levels and heat-treatments were also studied here.

\section{Measurement}
All coatings were produced by CSIRO (Commonwealth Scientific and Industrial Research Organization) and deposited upon both 1-inch silica discs and the silicon cantilevers commonly used in mechanical-loss measurements \cite{Martin09b}. The pure tantala samples were prepared as part of one coating run, and were heat-treated to 300, 400, 600, and 800 $^{\circ}$C. It was previously shown that the 800 $^{\circ}$C sample had begun to crystallize \cite{Martin10}. The titania-doped samples were made during a separate coating run and were either left untreated or heat-treated to 300, 400, or 600 $^{\circ}$C. The untreated samples are often referred to as As Deposited (AD); the deposition temperature was $\sim$100 $^{\circ}$C, so for the purpose of this paper, the results are considered as heat-treated to this temperature. Additional heat-treatment was carried out by annealing samples in air for 24 hours. All of the titania-doped tantala samples were found to be amorphous. Titania-doped samples were either 25 or 55\% doped, as measured by metal cation. All coatings were measured using ellipsometry to be $\sim$500 nm thick. 

\subsection{Nanoindentation}\label{sec:NanoIndent}
\begin{figure}
\begin{center}
\includegraphics[width=8.4cm]{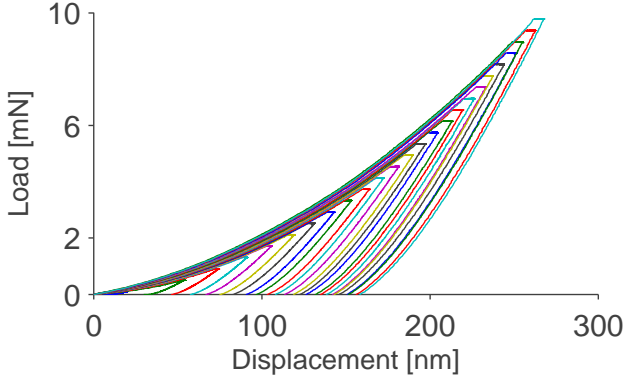}
\caption{Example of the Load-Displacement curves used in calculating the Young's modulus at one position on a film of pure tantala deposited on a silica substrate and heat-treated to 600 $^\circ$C. \label{fig:LoadDisplacement}}
\end{center}
\end{figure}

Nanoindentation is a technique developed to measure the mechanical properties of small volumes of materials in a simple fashion \cite{Mencik97}. These properties are measured by making indentations at the nanometer scale and recording the load, $P$, and displacement, $h$, response as the indenter is driven into and withdrawn from the material. An example of the nanoindentation load-displacement curves taken into tantala can be seen in Figure \ref{fig:LoadDisplacement}. In this example, the indents are made using the standard load-hold-unload cycle. A load is applied to the indentation tip, forcing it into the sample and increasing the displacement. During the loading phase, work is done as the sample is both elastically and plastically deformed. During the hold phase, the force is held constant, but the sample continues to deform due to creep effects which arise due to the movement of the material within the specimen under high pressure. During the unloading phase, the load is reduced and the indentation tip is withdrawn from the sample. This phase is characterized by having only an elastic response. We therefore analyze the unloading phase in order to measure the Young's modulus of the material. 

Using this technique, indentation
 measurements were made using the Hysitron TI-700 Ubi, located at the University of Cambridge, with a diamond Berkovich pyramidal tip. Effective moduli, composed of the coating and substrate Young's moduli, were extracted from the indentation data using the method of Oliver and Pharr \cite{Oliver04}. Once the load-displacement data is recorded, the elastic modulus is determined from:
\begin{equation}
\label{eqn:OPEstar}
E^* = \frac{\sqrt{\pi}}{2} \frac{d\!P}{d\!h} \frac{1}{\sqrt{A}},
\end{equation}
where $A$ is the projected area of contact under load, $d\!P/d\!h$ is the slope of the load-displacement curve at the beginning of the unloading phase, and $E^*$ is the combined modulus of the sample and indenter:
\begin{equation}
\frac{1}{E^*} = \frac{1-\nu_{\rm{i}}^2}{E_{\rm{i}}} + \frac{1-\nu_{\rm{s}}^2}{E_{\rm{s}}}.
\end{equation} 
Here, $E$ is the Young's modulus, and $\nu$ is the Poisson's ratio of the sample and indenter, marked with subscripts `s' and `i', respectively. 

The value of $d\!P/d\!h$ is generally extracted from the data by fitting an empirically-derived equation,
\begin{equation}
\label{eqn:OPExponential}
P = \alpha(h - h_{\rm{f}})^m,
\end{equation}
to the unloading portion of the curve. Here, $\alpha$ and $m$ are the fitting constants, and $h_{\rm{f}}$ is the displacement at zero load on the unloading curve. Once fit, the derivative of the P-h relation is taken at the maximum value of h, $h_{\rm{max}}$, to give the value of $d\!P/d\!h$. 

The area function is calculated by making a series of indents into a fused quartz reference sample with a known hardness and Young's modulus, solving Equation \ref{eqn:OPEstar} for $A$, and fitting the following equation \cite{Oliver92}:
\begin{equation}
A(h_{\rm{c}}) = C_1 h_{\rm{c}}^2 + C_2 h_{\rm{c}} + C_2 h_{\rm{c}}^{1/2} + C_4 h_{\rm{c}}^{1/4} + \ldots
\end{equation}
Here, $C_1$ is usually a number close to 24.5 for a Berkovich indenter, and the remaining constants are fit to indents made into samples of known modulus at different depths in order to account for tip rounding.

For a thin coating on much thicker substrate, if the Young's moduli of the coating and substrate differ, the modulus measured using the Oliver and Pharr method will vary with indentation depth \cite{Saha02}. This is due to the increasing influence of the substrate as the load is increased. In order to minimise this influence, it is often suggested that indents be made such that $h_{\rm{max}}$ is less than 10\% of the thickness of the coating, $t_c$ \cite{Oliver92}.  While this is generally acceptable for coatings greater than about a micron, it is not practical on thinner coatings where the errors in the area function and surface defects begin to have an effect at very small indentation depths \cite{FischerCripps06}.

\begin{figure}
\begin{center}
\includegraphics[width=8.4cm]{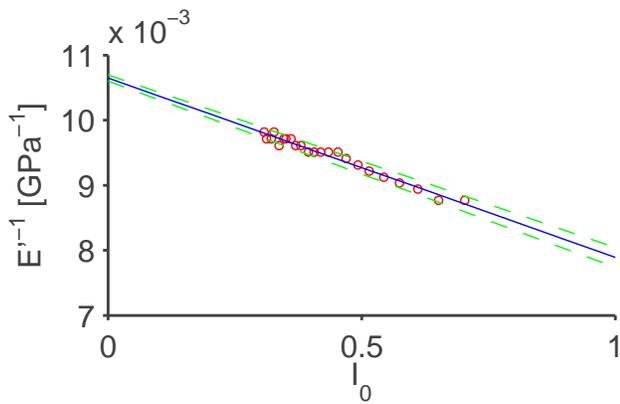}
\caption{Example of the fit of equation \ref{eqn:SongModel} to indentation data. The red circles are the values of $E'^{-1}$ from the Oliver and Pharr analysis against $I_0$, the blue line is the fit of the Song and Pharr model to the data, and the green dashed lines show the one standard deviation uncertainties to the fit. These data are from one position on the pure tantala sample heat treated to $400^{\circ}$ C. \label{fig:InvE_I0}}
\end{center}
\end{figure}

The coating modulus was extracted from the values taken from the Oliver and Pharr method using the model of Song and Pharr \cite{Song99,Rar01}. In the Song and Pharr model, the elastic moduli of the film and substrate are added in series, and weighted by a factor dependent upon the indentation area:
\begin{equation}
\label{eqn:SongModel}
\frac{1}{E'} = \frac{1}{E_{\rm{s}}} + \left(\frac{1}{E_{\rm{c}}} - \frac{1}{E_{\rm{s}}}\right)I_0(t/a).
\end{equation}
Here, the subscripts $c$ and $s$ represent the coating and substrate, respectively, and $I_0(t/a)$ is a weighing function that is equal to 1 for shallow indents and 0 for deep indents, and is given by the equation \cite{Gao92}:
\begin{widetext}
\begin{equation}
\label{eqn:Izero}
I_0(t/a) = \frac{2}{\pi} \arctan(t/a) + \frac{1}{2 \pi (1-\nu)} \left[(1-2\nu) (t/a) \ln\left(\frac{1+(t/a)^2}{(t/a)^2}\right) - \frac{t/a}{1+(t/a)^2}\right],
\end{equation}  
\end{widetext}
where $a$ is the radius of a circle with the equivalent area as the projected area of indent, $\pi a^2 = A(h_{\rm{c}})$, and $t$ is the difference between the thickness of the coating and the contact depth of the indent \cite{Rar01}. A plot of $E'^{-1}$ against $I_0$ for a number of indents made at different depths will yield a linear relationship with y-intercept of $E_{\rm{s}}^{-1}$ and a value of $E_{\rm{c}}^{-1}$ at \mbox{$I_0 = 1$}. An example of this can be seen in Figure \ref{fig:InvE_I0}, showing $E'^{-1}$ as a function of $I_0$ measured on one of the pure tantala samples, heat-treated to $400^{\circ}$ C, in red circles, and the fit of equation \ref{eqn:SongModel} in blue. The green dotted lines are the one standard deviation uncertainties from the fit, assuming that the noise in the data are Gaussian.

This method requires indents at varying depth, so for each location on a sample, at least 25 indents were made, varying the maximum applied load evenly between 500 and 10000 $\rm{\mu}$N. At least two positions were measured on each sample, and the weighted mean of the extracted coating moduli was calculated and is given in Table \ref{tab:IndentResults}.

For the pure tantala samples, indentations were made into coatings deposited onto both silicon and silica substrates. Finite element analysis suggests that indents into coatings on the more compliant silica substrates will have an additional substrate effect that would artificially reduce the extracted coating modulus by as much as 5\% \cite{Hay11}. Comparing the moduli measured on silica to those measured on silicon, the moduli measured on the silica substrate samples are in fact $\sim$5\% lower. This uncertainty is included as a +5\% systematic uncertainty for all the samples on silica substrates in the table. The modulus extracted using the Song and Pharr method also depends upon the Poisson ratio of the coating material. The Poisson ratios of amorphous tantala and titania-doped tantala are unknown, however the Poisson ratios of similar amorphous metal oxides tend to be in the range of 0.20-0.30. The values listed in the table are given using an assumed value for the Poisson ratio of 0.25. Variation within the range of 0.20-0.30 varies the resulting moduli by less than 3\% in all cases.   

\begin{table*}
%\centering
\caption{Young's moduli measured for various titania-doped tantala films. Included are the temperatures of heat-treatment for the samples, the number of positions measured on each sample, the Young's moduli measured, along with the statistical uncertainty from the measured indents on each sample, and the systematic uncertainty which arises from the softer substrate and uncertainty in the Poisson ratio of the coating materials.\label{tab:IndentResults}}
%\begin{ruledtabular}
\begin{tabular}{|c|c|c|c|c|c|c|}
\hline \hline
\% Ti&H-T [$^\circ$ C]&Substrate&Pos. Meas.&E [GPa]&Stat. Uncert. [GPa]&Sys. Uncert. \%\\
\hline
0&300&SiO2&5&152&2.6&$\pm$3+5\\
0&400&SiO2&5&137&1.1&$\pm$3+5\\
0&600&SiO2&5&133&1.2&$\pm$3+5\\
0&800&SiO2&5&162&6.4&$\pm$3+5\\
0&300&Si&2&160&14.2&$\pm$3\\
0&400&Si&2&146&3.3&$\pm$3\\
0&600&Si&2&137&3.7&$\pm$3\\
\hline
25&AD&SiO2&3&143&2.6&$\pm$3+5\\
25&300&SiO2&2&137&1.7&$\pm$3+5\\
25&400&SiO2&3&145&3.2&$\pm$3+5\\
25&600&SiO2&7&132&1.1&$\pm$3+5\\
\hline
55&AD&SiO2&3&145&4.9&$\pm$3+5\\
55&300&SiO2&3&158&2.9&$\pm$3+5\\
55&400&SiO2&3&142&1.7&$\pm$3+5\\
55&600&SiO2&4&177&4.3&$\pm$3+5\\
\hline \hline
\end{tabular}
\end{table*}

\subsection{Thermal Bending}
Measurements of the CTE were made using silicon cantilevers of dimensions 34 mm long x 5 mm wide and $\sim$115 $\mu$m thick, similar to those used in cryogenic mechanical loss measurements \cite{Abernathy11,Martin10,Martin09b,Reid06a}. The CTE of a coating deposited upon such a cantilever can be determined by measuring the variation in coating stress with sample temperature. Following the modified Stoney's formula \cite{Stoney09,Hoffman66}, 
\begin{equation}
\label{eqn:BendingStoney}
\sigma_{\rm{coating}} = \frac{1}{6} B_{\rm{s}} \frac{t_{\rm{s}}^{2}}{t_{\rm{c}}} \left(\frac{1}{R_{0}} - \frac{1}{R} \right),
\end{equation}
the change in the radius of curvature of a coated substrate, $R$, from that of the uncoated substrate, $R_{0}$, is related to the stress in the coating $\sigma_{\rm{coating}}$, where $B_{\rm{s}}$ is the biaxial modulus of the substrate, $t_{\rm{s}}$ is the thickness of the substrate, and $t_{\rm{c}}$ is the thickness of the coating. The biaxial modulus of a material is defined as: $B = E/(1-\nu)$. Some of this stress is related to the thermal expansion mismatch, $\alpha_{\rm{c}} - \alpha_{\rm{s}}$, between the coating and the substrate, 
\begin{equation}
\label{eqn:BendingStressTemperature} 
\sigma_{\rm{coating}} = \sigma_{\rm{I}} +(\alpha_{\rm{c}} - \alpha_{\rm{s}}) B_{\rm{c}} \Delta\!T,
\end{equation}
where $\sigma_{\rm{I}}$ is the intrinsic stress in the coating from non-thermal sources, and $\Delta\!T$ is the temperature difference from the last significant thermal treatment of the coating, such as deposition or heat-treatment \cite{Doerner88}. Therefore, the variation in the stress as a function of temperature yields the following relation:
\begin{equation}\label{eqn:StressSlope}
d\!\sigma_{\rm{coating}}/d\!T = (\alpha_{\rm{s}} - \alpha_{\rm{c}})B_{\rm{c}}.
\end{equation}
%where $\sigma$ is the coating stress, $T$ is the sample temperature, $\alpha$ is the thermal expansion coefficient of the coating, subscript c, and the substrate, subscript s, and $B_{\rm{c}}$ is the biaxial modulus of the coating: $B = Y/(1-\nu)$ where $\nu$ is the Poisson ratio.

\begin{figure}
\begin{center}
\includegraphics[width=8.4cm]{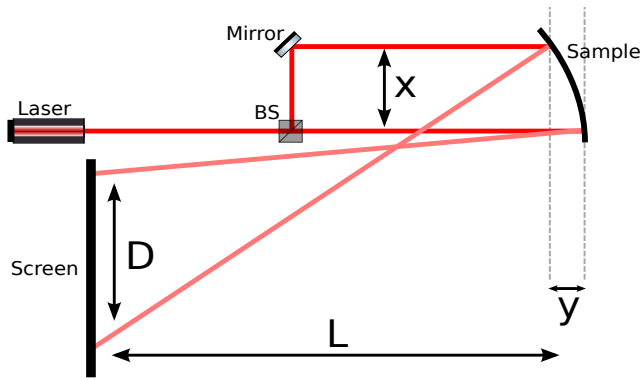}
\caption{Diagram of the thermal bending experimental setup. The measurement of the spot separation, $D$, is dependent upon the displacement of the cantilever tip, $y$. This diagram is not to scale.\label{fig:BendingSketch}}
\end{center}
\end{figure}

A simple apparatus was designed in order to measure the radius of curvature of coated cantilever samples. In this apparatus, shown schematically in Figure \ref{fig:BendingSketch}, a laser beam is separated into two parallel beams using a beam-splitter and $45^{\circ}$ mirror, separated by a distance, $x$. These beams are reflected from the cantilever, with one spot reflecting from very near the clamped base of the cantilever, and the other spot reflecting from the tip. These two reflected beams are incident upon a screen placed a distance $L$ from the sample. 

The distance between the spots on the screen, $D$, will be the sum of the original separation of the beams and the deviation caused by the curvature of the cantilever: $D = x + \delta$, where $D$ is negative if the beams cross between the sample and the screen. Therefore, if $D$ is negative or less than $x$, the beams are convergent, and the sample is concave (as drawn in Figure \ref{fig:BendingSketch}); if $D$ is greater than $x$, then the beams are divergent, and the sample is convex. If the displacement of the sample tip, $y$ is small relative to $x$, the radius of curvature of the sample can be calculated using the relation:
\begin{equation}
\label{eqn:BendingRGeometry}
R = 2 L x / \delta.
\end{equation} 
This device was able to measure the radius of curvature of the samples to an accuracy of $\sim$5\%, as tested in measurements of concave mirrors of known radius. In order to make measurements at varying temperatures, the samples were placed within an insulated copper box with a thin transparent opening to allow the passage of the laser beams. The temperature within the box was controlled using resistive heaters mounted inside the box and the temperature was measured using a thermocouple mounted within the cantilever clamp. 

\begin{figure}
\begin{center}
\includegraphics[width=8.4cm]{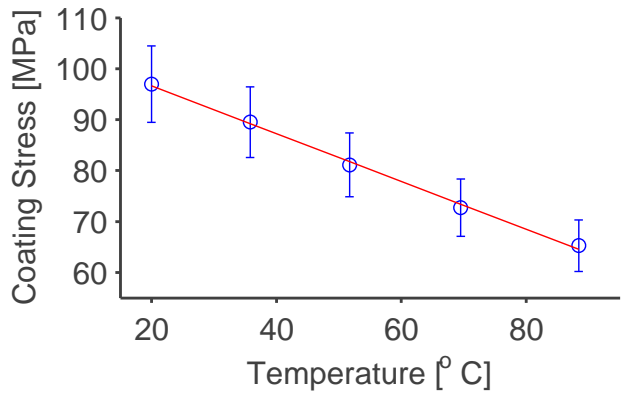}
\caption{Plot of tensile stress (blue circles) versus temperature with accompanying fitted line (red). The slope of this line is used to calculate the thermal expansion coefficient of the coating material.\label{fig:SigmaT}}
\end{center}
\end{figure}

Measurements of the radius of curvature were made at intervals between 25 and 100 $^\circ$C. The radius was converted to stress, and the stress variation with temperature was fit with a straight line to give the slope in Equation \ref{eqn:StressSlope}. An example of the resulting plot of stress as a function of temperature is given in Figure \ref{fig:SigmaT}. In this plot, the error bars are from the statistical uncertainty in the measurements as well as the systematic uncertainty in all the components of equation \ref{eqn:BendingStoney} except for the thickness of the substrate. Our uncertainty in the thickness of the substrate can add a systematic error of as much as 20\% to the measurements of tensile stress; however, even these uncertainties are utilized in the calculation of the thermal expansion. By taking the Young's modulus measured using the nanoindention procedure discussed above, and assuming a Poisson ratio of $0.25\pm0.05$, the coefficient of thermal expansion could be calculated for each sample. The results are shown in Figure \ref{fig:ThermalExpansion}.

\section{Results}
%\begin{figure}[t]
%        \centering
%        \begin{subfigure}[t]{0.5\textwidth}
%                \centering
%                \includegraphics[width=\textwidth]{{AllTantalaResults}}
%                \caption{Mean Young's moduli of all tantala samples measured on silica substrates, plotted for samples of different heat-treatment. The sample heat treated at 800$^\circ$ C was found to be poly-crystalline, all others are amorphous.}
%                \label{fig:AllTantalaResults}
%        \end{subfigure}%
%        ~ %add desired spacing between images, e. g. ~, \quad, \qquad etc.
%          %(or a blank line to force the subfigure onto a new line)
%        \begin{subfigure}[t]{0.5\textwidth}
%                \centering
%                \includegraphics[width=\textwidth]{ThermalExpansion}
%                \caption{Coefficient of thermal expansion measured using the thermal bending technique on titania-doped tantala samples heat-treated at different temperatures. The solid lines indicate the weighted mean of each set.}

%		\label{fig:ThermalExpansion}
%        \end{subfigure}
%\end{figure}

\subsection{Young's Modulus}
\begin{figure}
\begin{center}
\includegraphics[width=8.4cm]{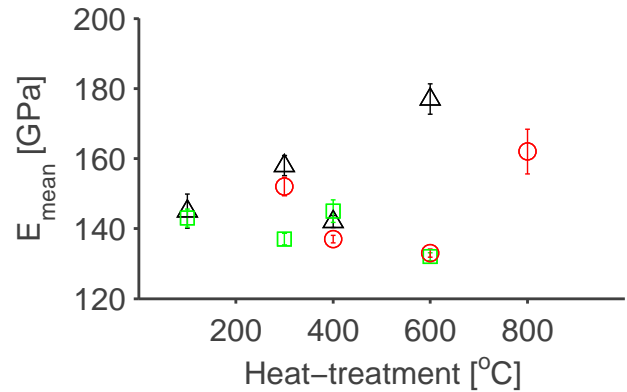}
\caption{Mean Young's moduli of all tantala samples measured on silica substrates, plotted for samples of different heat-treatment. The red circles are pure tantala, the green squares are 25\% titania-doped tantala, and the black triangles are 55\% titania-doped tantala. Error bars are the statistical uncertainty on the weighted mean. The sample heat treated at 800$^\circ$ C was found to be poly-crystalline, all others are amorphous.\label{fig:AllTantalaResults}}
\end{center}
\end{figure}

Coating Young's moduli as measured on silica substrates are shown in Figure \ref{fig:AllTantalaResults} as a function of heat-treatment. Looking only at the pure tantala (red circles), the Young's modulus appears to decrease with increasing heat-treatment until the coating begins to crystallize between 600 and 800 $^\circ$C. Such a trend was postulated in \cite{Brown04}, as an indicator of increasing void space with increased heat-treatment. A similar trend is seen with the 25\% titania-doped samples (green squares), with the exception of the 400 $^\circ$C sample. The opposite trend is seen with the 55\% titania-doped samples (black, upward triangles); again, with the exception of the 400$^\circ$ C sample. This is most likely due to the abundance of titania, which is known to have a low crystallization temperature and a high Young's modulus \cite{Gaillard09,Kurosaki05}. The two titania-doped samples heat-treated at 400$^\circ$ C were produced at the same time, they were most likely heat-treated together, and may not have been fully heat-treated; this agrees with the fact that the two samples give approximately the same moduli as the As Deposited coatings. Overall, it is shown that while the commonly measured value of $\sim$140 GPa \cite{Cetinorgu09,KlembergSapieha04} is a reasonable value for the Young's modulus of tantala, it is dependent upon both the doping and heat-treatment of the coating.

\subsection{Thermal Expansion}  
\begin{figure}
\begin{center}
\includegraphics[width=8.4cm]{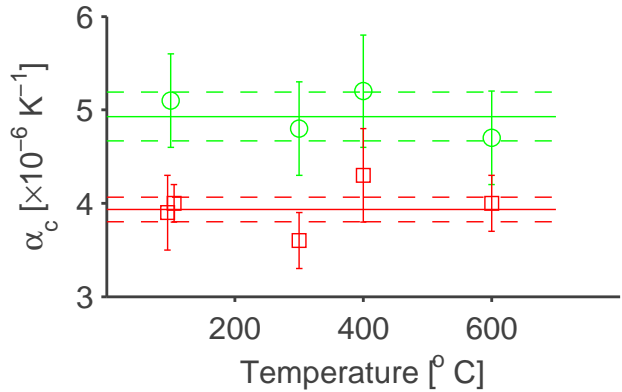}
\caption{Coefficient of thermal expansion measured using the thermal bending technique for 55\% (green circles) and 25\% (red squares) titania-doped tantala samples heat-treated at different temperatures. The solid lines indicate the weighted mean of each set.\label{fig:ThermalExpansion}}
\end{center}
\end{figure}

The coefficients of thermal expansion for both of the titania doping concentrations are shown in Figure \ref{fig:ThermalExpansion}. The plot suggests that there is no obvious trend in the coefficient of thermal expansion of the coatings with heat-treatment temperature. Also plotted as solid lines are the weighted means of the coefficients. For the 55\% titania-doped tantala, the mean coefficient of thermal expansion is $(4.9\pm0.3) \times10^{-6} \text{ K}^{-1}$, and for the 25\% titania-doped samples, the mean is $(3.9\pm0.1) \times10^{-6} \text{ K}^{-1}$. These values can be compared to the reported coefficient of thermal expansion of pure tantala coatings deposited by dual ion-beam sputtering in \cite{Cetinorgu09}, which gives a value of $4.4\times10^{-6}\rm{K}^{-1}$. Both of these values are higher than that used for pure tantala, however, they are probably affected by the presence of the titania doping. Unfortunately, we were unable to measure any samples of pure tantala.  %Measurements of the coefficient of thermal expansion of bulk tantala give values of $6.7\times10^{-6}\rm{ K}^{-1}$ for sintered bars \cite{Wu06}, and $3.0\times10^{-6}\rm{K}^{-1}$ for chemically vapour deposited $\beta$-Ta$_2$O$_5$ \cite{Bae95}, and $2.06\times10^{-6}\rm{K}^{-1}$ and $2.45\times10^{-6}\rm{K}^{-1}$ for $\alpha$- and $\beta$-Ta$_2$O$_5$ respectively \cite{Touloukian70}. The values measured here are of similar or moderately greater values than those in the literature, which may be due to the effects of the titania dopant.

\section{Conclusion}
We have made direct measurements of the coefficient of thermal expansion and Young's modulus of pure and titania-doped tantala heat-treated at various temperatures. Our results indicate that the Young's modulus of IBS tantala films is affected by both post-deposition heat-treatment and titania doping, and that the coefficient of thermal expansion is not affected by heat-treatment, but is altered by titania doping. We have also made direct measurements of coating materials very similar to those used in second generation interferometric gravitational wave detectors, confirming previous estimates and allowing for greater confidence in the prediction of thermal noise levels within these detectors. 

%To date, the values used by the gravitational wave community for the coefficient of thermal expansion and Young's modulus of pure and titania-doped tantala have been $3.6\times10^{-6}\rm{ K}^{-1}$ and 140 GPa, respectively. The advanced LIGO detector, currently under commision, will utilize mirrors coated with $\sim$25\% titania-doped tantala/silica mirror coatings optimized to reduce coating Brownian noise and heat-treated to 500-600$^\circ$ C. This paper serves to verify these numbers, as they are not far removed from those measured here. 
\begin{acknowledgments}
The authors would like to acknowledge the contributions of the Oyen Group from the Cambridge University Department of Engineering, especially Oliver Hudson, Tamaryn Shean, and Daniel Strange, for their aid in obtaining the nanoindentation data.

IWM holds a Royal Society University Research Fellowship and SR is a Royal Society-Wolfson Research Merit Award holder. The authors would like to thank the UK Science and Technology Facilities Council, the University of Glasgow, the Scottish Universities Physics Alliance and the Scottish Founding Council for financial support. We also wish to thank our colleagues in the GEO600 and LIGO Scientific Collaboration for their interest in this work.
 
The LIGO Observatories were constructed by the California Institute of Technology and
Massachusetts Institute of Technology with funding from the National Science Foundation
under cooperative agreement PHY-9210038. The LIGO Laboratory operates under
cooperative agreement PHY-0107417. This paper has been assigned LIGO Document Number LIGO-
P1300107.

\end{acknowledgments}

\end{document}